\documentclass{elsart}

\usepackage{natbib}

\usepackage{amssymb}

\def\beq{\begin{equation}}
\def\eeq{\end{equation}}
\def\_#1{_{\scriptscriptstyle #1}}
\def\^#1{^{\scriptscriptstyle #1}}
\def\vg{\vec g}
\def\vu{\vec u}
\def\vr{\vec r}
\def\vv{\vec v}
\def\va{\vec a}
\def\vx{\vec x}
\def\Q{\vec Q}
\def\vg{{\bf g}}
\def\vu{{\bf u}}
\def\vr{{\bf r}}
\def\vv{{\bf v}}
\def\va{{\bf a}}
\def\vs{{\bf s}}
\def\vF{{\bf F}}
\def\vA{{\bf A}}
\def\vx{{\bf x}}
\def\Q{{\bf Q}}
\def\grad{\vec\nabla}
\def\div{\vec\nabla\cdot}
\def\ud#1{{}_{,#1}}

\def\a0{$a_0$}
\def\grad{\vec\nabla}
\def\gf{\grad \phi}
\def\gfs{(\grad \phi)^2}
\def\ff{\varphi}
\def\gff{\grad \ff}
\def\gffs{(\grad \ff)^2}
\def\tf{\tilde\ff}
\begin{document}

\begin{frontmatter}

% Title, authors and addresses

% use the thanksref command within \title, \author or \address for footnotes;
% use the corauthref command within \author for corresponding author footnotes;
% use the ead command for the email address,
% and the form \ead[url] for the home page:
% \title{Title\thanksref{label1}}
% \thanks[label1]{}
% \author{Name\corauthref{cor1}\thanksref{label2}}
% \ead{email address}
% \ead[url]{home page}
% \thanks[label2]{}
% \corauth[cor1]{}
% \address{Address\thanksref{label3}}
% \thanks[label3]{}

\title{MOND--theoretical aspects}

% use optional labels to link authors explicitly to addresses:
% \author[label1,label2]{}
% \address[label1]{}
% \address[label2]{}

\author{Mordehai Milgrom}

\address{Department of Condensed Matter Physics, Weizmann
Institute, Rehovot, Israel}

\begin{abstract}
% Text of abstract
I discuss open theoretical questions pertaining to the modified
dynamics (MOND)--a proposed alternative to dark matter, which
posits a breakdown of Newtonian dynamics in the limit of small
accelerations. In particular, I point the reasons for thinking
that MOND is an effective theory--perhaps, despite appearance, not
even in conflict with GR. I then contrast the two interpretations
of MOND as modified gravity and as modified inertia. I describe
two mechanical models that are described by potential theories
similar to (non-relativistic) MOND: a potential-flow model, and a
membrane model. These might shed some light on a possible origin
of MOND. The possible involvement of vacuum effects is also
speculated on.
\end{abstract}

\begin{keyword}
% keywords here, in the form: keyword \sep keyword
Cosmology \sep the dark matter problem \sep galaxy dynamics
% PACS codes here, in the form: \PACS code \sep code

\end{keyword}

\end{frontmatter}

% main text
\section{Introduction--the modified dynamics}
MOND is a modification of Newtonian dynamics in a form that
obviates the need for dark matter, when
 applied to galactic systems.
It does this by introducing a constant with the dimensions of
 an acceleration,
\a0, and positing that standard Newtonian dynamics is a good
approximation only for accelerations that are much larger than
\a0. The exact behavior in the opposite limit is described by
specific underlying theories like those described below. However,
the basic point of MOND, from which follow most of the main
predictions, can be encapsuled in the following approximate
relation:
 a test particle at a distance $r$ from a mass $M$ is
 subject to the acceleration $a$ given by
 \beq a^2/a_0\approx MGr^{-2}, \label{basic} \eeq
  when
 $a\ll a_0$, instead of the standard expression $a=MGr^{-2}$,
 which holds when $a\gg a_0$. Or, somewhat more generally, if
 $a_N$ is the Newtonian expression for the acceleration, then
 for $a_N\ll a_0$
\beq a\approx (a_Na_0)^{1/2}, \label{simpla} \eeq
instead of $a\approx a_N$, which holds for $a_N\gg a_0$.
 The two expressions may be interpolated to give the heuristic
 relation
 \beq \mu(a/a_0)a\approx a_N, \label{mond} \eeq
 where the interpolating function $\mu(x)$ satisfies
 $\mu(x)\approx 1$ when $x\gg 1$, and $\mu(x)\approx x$ when $x\ll
 1$. This expression, while lacking from the formal point of view,
 is very transparent, and captures the essence of MOND. I shall
 describe below more presentable theories based on this basic
 relation, but these are still phenomenological theories into
 which the form of $\mu(x)$ has to be put in by hand. It will
 hopefully follow one day from a more basic
 underlying theory for MOND, which we still lack. Most of the
 implications of MOND do not depend strongly on
the exact form of $\mu$. Much of the phenomenology pertinent to
the mass discrepancy in galactic systems occurs in the deep-MOND
regime ($a\ll a_0$), anyway, where we know that $\mu(x)\approx x$.

\section{MOND phenomenology}
The phenomenology dictated by MOND, and its application and
testing in galactic systems: galaxies of all sorts, galaxy groups,
clusters, and super-clusters, is discussed in recent reviews; e.g.
\citet{mildark98} and \citet{araa}. Here I only touch briefly on a
few general aspects of the phenomenology.
\par
It is important to recognize the message that the phenomenological
success of MOND would carry. A sentiment is expressed occasionally
that MOND--successful as it may be--is only a hypothesis that
``saves the phenomena''; that MOND phenomenology might one day be
understood within the dark-matter doctrine [e.g., \cite{kt}, but
see \cite{milkt}]. To be sure, this is still a far cry; and to
appreciate how tall an order it would be, we note that the MOND
idea, if it is taken just as the distillation of DM phenomenology,
entails not one, but many {\it independent} laws that govern the
mass discrepancy in galactic systems: the analog of Kepler laws in
the solar system. Some of these laws involve \a0, and some do not.
In those that do, \a0 appears in several independent roles if
viewed just as a phenomenological parameter. (In the framework of
MOND \a0 appears in two roles: the borderline acceleration between
the MOND and the Newtonian regime, and a bench-mark acceleration
deep in the MOND regime.)

 Here are some of the laws:

1. Galaxy rotation curves are asymptotically flat.

2. The asymptotic rotational velocity of a galaxy is proportional
to the fourth root of its total baryonic {\it mass}: the baryonic
Tully-Fisher relation. The proportionality constant is
$(Ga_0)^{1/4}$ (the role of \a0 as a deep-MOND parameter).

3. Galaxies with high central surface mass
densities--corresponding to accelerations larger than \a0--should
show no mass discrepancy in the inner parts. The discrepancy
should appear only beyond a certain radius but always when the
acceleration becomes comparable with \a0. (This is \a0 in its role
as borderline acceleration.)

4. In LSB galaxies, which have low surface density everywhere, the
mass discrepancy should start right from the center, and its
magnitude is given by the inverse of the acceleration in units of
\a0 (\a0 as a deep-MOND fiducial acceleration).

5. Self-gravitating, quasi-isothermal spheres cannot have mean
accelerations much exceeding \a0 (or mean surface densities much
exceeding $\sim a_0G^{-1}$)--with implications for round galactic
systems, e.g. underlying the Fish law for elliptical galaxies (\a0
as a borderline acceleration).

Laws 1-4 are special cases of the sweeping law:

6. The distribution of the visible (baryonic) matter in every
galaxy tightly determines the acceleration field of a galaxy and
thus  the distribution of the DM. The relation between the two
mass distributions is given by eq.~(\ref{mond}), where \a0 appears
in both roles.
 This should hold despite the very different
 formation-evolution-interaction histories that the different
 galaxies have undergone.

Analogous laws hold for galactic systems other than galaxies. And
they should be counted as separate laws in the framework of the DM
picture. The fact that these systems look so different on the sky
tells us that their baryonic component, at least, have undergone
very different histories, so there is no reason, for example, why
law 6 above would carry from one system type to another.
\par
 The above
laws are phenomenologically independent in the sense that in the
framework of the dark matter paradigm it is easy to conceive of
baryon-plus-dark-matter galactic systems that obey any set of
these laws, while breaking the others \citep[see a more detailed
discussion with examples in][]{milkt}. This is similar, for
example, to the existence of different quantum phenomena; e.g.,
the black-body spectrum,  the photoelectric effect, the hydrogen
spectrum, superconductivity, etc., in which the Planck constant
appears in different roles. Without the quantum theory, these all
seem unrelated. MOND is, likewise, a theory that unifies all the
above laws of galaxy dynamics phenomenology.
\par
The strength of the case for MOND is further augmented by noting
that, historically, these laws were not arrived at by distillation
of existing data. They were, most of them, pure predictions of
 MOND. The only input in the construction of MOND was the
hypothesized asymptotic velocity behavior in disc galaxies, to
wit, laws 1 and 2 above taken as axioms, and assumed only for disc
galaxies. And even these two were not established at the time to
the extent they are today.

\par
The second general point I want to make concerns a  possibly very
significant coincidence involving the actual value of \a0. The
value
 that fits the data discussed above is about $10^{-8}{\rm cm}~{\rm s}^{-2}$. This
value of \a0 is of the order of some acceleration constants of
cosmological significance. It is very nearly
 $a_{ex}\equiv c\bar H$, where $\bar H\equiv H_0/2\pi$  ($H_0$
  is the Hubble constant) and, it is
 also of the order of $a_{cc}\equiv c(\Lambda/3)^{1/2}$,
 where $\Lambda$ is the emerging value of the cosmological
 constant (or ``dark energy'').
\par
Because the cosmological state of the universe changes, such a
connection, if it is a lasting one, may imply that galaxy
evolution does not occur in isolation, affected only by nearby
objects, but is, in fact, responding constantly to changes in the
state of the universe at large. For example, if the connection of
$a_0$ with the Hubble constant always holds, the changing of the
Hubble constant would imply that $a_0$ must change over cosmic
times, and with it the appearance of galactic systems, whose
dynamics $a_0$ controls. If, on the other hand, \a0 is a
reflection of a true cosmological constant, then is might be a
veritable constant.

\section{MOND as an effective theory}
The above proximity of \a0 to the cosmological acceleration scale,
beyond its phenomenological significance, may hint at a deep
connection between cosmology and local dynamics in systems that
are very small on cosmological scales. Either cosmology somehow
enters and affects local laws of physics, such as the law of
inertia or the law of gravity, or a common agent affects both
cosmology and local physics so as to leave on them the same
imprint. This would mean that MOND--and perhaps more cherished
notions, such as inertia--is a derived concept: an effective
theory. An observed relation between seemingly unrelated constants
appearing in a theory (in our case, \a0, the speed of light, and
the radius of the horizon) may indicate that MOND is only an
approximation of a theory at a deeper stratum, in which some of
the constants do not really have any special role. For example, in
experiments and observations confined to the vicinity of the
earth's surface, the free-fall acceleration, $g$ attains the
status of a ``constant of nature''. It is numerically related to
two other important ``constants'': the escape speed $c_e$ (objects
thrown with a higher velocity never return) and the radius of the
earth $R_\oplus$. (This relation, $g=c_e^2/2R_\oplus$, is
practically the same as that between $a_0$, the speed of light,
and the Hubble radius, in MOND.) But looking beyond the surface,
and knowing about universal gravity, we know that all these
``constants'' actually derive from the mass and radius of the
earth (hence the relation between the three). They are useful
parameters when describing near surface phenomena, but quite
useless in most other circumstances. In a similar vein, $a_0$
might turn out to be a derived constant, perhaps variable on
cosmic time scales, perhaps even of no significance beyond the
non-relativistic regime, where MOND has been applied so far. Its
connection with the speed of light and the radius of the universe
will, hopefully, follow naturally in the underlying theory that
still eludes us.
\par
Many instances of such effective theories are known. Even General
Relativity is now thought to be an effective, low-energy
approximation of a higher theory (e.g. a string-inspired theory);
an idea that has been anticipated by Sakharov's ``induced
gravity'' idea.

\section{Interpretations}
Equations(\ref{basic})-(\ref{mond})  have the form of a
modification of the law of inertia, but since they are algebraic
relations between the MOND and Newtonian accelerations they can
simply be inverted to read $a=F/m=a_Nf(a_N/a_0)$, which seems to
leave the second law intact, while modifying the Newtonian
gravitational force $ma_N$ to the MOND value $ma$. Because
gravitation is the sole force that governs galactic dynamics--the
only corner where the mass discrepancy has been clearly
observed--existing phenomenology does not distinguish well between
the interpretations of MOND as modified gravity, and modified
inertia. Although there are matter-of-principle differences
between the two interpretations (see below) they pertain to
observations that are not yet available. For now we must then
investigate both options.
\par
But what exactly is meant by ``modifying gravity'', and
``modifying inertia''? When dealing with pure gravity the
distinction is not always clear. For example, the Brans-Dicke
theory may be viewed as either. But when other interactions are
involved, the distinction is clear. Obviously, modified inertia
will enter the dynamics of systems even when gravity is
negligible, unlike the case for modified gravity. Formally, the
distinction might be made as follows. In a theory governed by an
action principle we distinguish three part in the action: The pure
gravitational part (for example, the Einstein-Hilbert action in
GR), the free action of the matter degrees of freedom (in GR it
also encapsules their interaction with gravity), and the action of
interactions between matter degrees of freedom (in GR they too
engender sources for gravity). By ``modifying gravity'' I mean
modifying the pure-gravity action; by ``modifying inertia'' I mean
modifying the kinetic (free) matter actions.
\par
 To understand this definition remember that
 inertia is what endows the
motion of physical objects (particles, fields, large bodies, etc.)
with energy and momentum--a currency in the physical world. Motion
itself is only of a descriptive value; inertia puts a cost on it.
For each kind of object it tells us how much energy and momentum
we have to invest, or take away, to change its state of motion by
so much. This information is encapsuled in the kinetic action,
which encodes the energy-momentum of the free degrees of freedom.
\par
For example, take the standard, non-relativistic action for a
system of particles interacting through gravity.

\beq S=S_{\phi}+S_k+S_{in}=-(8\pi G)^{-1}\int d^3r~(\grad\phi)^2
+\sum_i(1/2)m_i\int dt~v_i^2-\int d^3r~\rho(\vr)\phi(\vr),
\label{acpar}\eeq
 where $\rho(\vr)=\sum_i
m_i\delta(\vr-\vr_i)$, and $m_i$, $r_i$ are the particle masses
and positions. In GR, $S_k$ and $S_{in}$ are lumped together into
the particle kinetic action $-\int d\tau$ (tau is the proper time
of the particle) .
\par
Here, modifying gravity would mean modifying $S_{\phi}$, while
modifying inertia would entail changing $S_k$.

\section{MOND as modified gravity}
An implementation of MOND as a non-relativistic modified gravity
was discussed by \cite{bm}, who replaced the standard, free,
potential action $S_{\phi}$ in eq.~(\ref{acpar})  by an action of
the form

\beq S_{\phi}=-(8\pi G)^{-1} a_0^2\int
d^3r~F[(\grad\phi)^2/a_0^2].\label{act} \eeq This gives, upon
variation on $\phi$, the equation

 \beq\div[\mu(|\grad\phi|)\grad\phi]=4\pi G\rho(\vr),
 \label{mondpoiss} \eeq
where $\mu(z)\equiv dF(y)/dy\vert_{y=z^2}$. This theory, since it
is derived from an action that has all the usual symmetries,
satisfies all the standard conservation laws. Its various
implications have been discussed in \cite{bm}, \cite{milsol},
\cite{milconformal}, and others.
\par
An important point to note is that this theory gives the desired
center-of-mass motion of composite systems: Stars, star clusters,
etc. moving in a galaxy with a low center-of-mass acceleration are
made of constituents whose internal accelerations are much higher
than \a0. If we look at individual constituents we see bodies
whose total accelerations are high and so whose overall motion is
very nearly Newtonian. Yet, their motion should somehow combine to
give a MOND motion for the center of mass. This is satisfied in
the above theory as shown in \cite{bm}. A similar situation exists
in GR. Imagine a tightly bound system of black holes, moving in
the weak field of a galaxy, say. While the motions of the
individual components are highly relativistic, and are governed by
a non-linear theory, we know that these motions combine to give a
simple Newtonian motion for the center of mass.
\par
This field equation, generically, requires numerical solution, but
it is straightforward to solve in cases of high symmetry
(spherical, cylindrical, or planar symmetry), where the
application of the Gauss law to eq.~(\ref{mondpoiss}) gives the
exact algebraic relation between the MOND ($\vg=-\grad\phi$) and
Newtonian ($\vg_N=-\grad\phi_N$) acceleration fields:

\beq \mu(g/a_0)\vg=\vg_N, \label{algebraic} \eeq which is
identical to the heuristic MOND relation we started with. Note
that in general, for configurations of lower symmetry,  this
algebraic relation does not hold (and, in general, $\vg$ and
$\vg_N$ are not even parallel).
\par
It is worth pointing out that in such a modified-gravity theory,
the deep-MOND limit corresponds to a theory that is conformally
invariant, as discussed in \cite{milconformal}. Whether this has
some fundamental bearings is not clear, but it does make MOND
unique, and enables one to derive useful analytic results, such as
an expression for the two-body force, and a virial relation,
despite the obstacle of nonlinearity.
\subsection{mechanistic models}
Inasmuch as MOND is still in need of an underlying theory it may
be useful to study mechanistic models or analogues that reproduce
similar phenomenology. These may help elucidate the origin of the
nonlinearity in (nonrelativistic) MOND, and perhaps the appearance
of the same acceleration constant in both local dynamics and
cosmology.

\par
 There is a large number of
physical phenomena that are governed by an equation like
eq.~(\ref{mondpoiss}), each with its own form of the function
$\mu(x)$, as detailed, e.g., in \cite{milconformal}, or
\cite{milnonlin}. By choosing the right underlying physics a form
of $\mu$ may perhaps be found that will correspond to MOND
behavior.

\par
It is well known, for example, that a stationary, potential flow
is described by the Poisson equation: If the velocity field
$\vu(\vr)$ is derived from a potential, $\vu=\grad\phi$, then the
continuity equation, which here determines the flow, reads
$\div\grad\phi=s(\vr)/\varrho_0$, where $s(\vr)$ is the source
density, and $\varrho_0$ is the (constant) density of the fluid.
When the fluid is compressible, but still irrotational, and
barotropic [i.e. has an equation of state of the form $p=p(\rho)$]
the stationary flow is described by the nonlinear Poisson
equation. The Euler equation reduces to Bernoulli's law

\beq h(\varrho)=-u^2/2+{\rm const.}, \label{bern} \eeq where
$dh/d\rho\equiv \rho^{-1}dp/d\rho$. This tell us that $\varrho$ is
a function of $u=|\grad\phi|$. Substituting this in the continuity
equation gives

\beq \div[\varrho(|\grad\phi|)\grad\phi]=s(\vr), \label{flow} \eeq
which has the same form as eq.~(\ref{mondpoiss}) if we identify
$\varrho$ as $\mu$, and the source density $s$ with the normalized
gravitational mass density $4\pi G\rho$ . Note, however, that from
the Bernoulli law, $d\varrho/d|u|= -\varrho|u|/c^2$, where
$c^2=dp/d\varrho$ is the formal squared speed of sound. Thus, in
the case of MOND, where $\mu$ is an increasing function of its
argument, the model fluid has to have a negative compressibility
$c^2<0$.  A cosmological-constant-like equation of state,
$p=-c^2\varrho$, with $c$ the speed of light gives
$\varrho(u)=\varrho_0 \exp(u^2/2c^2)$, which is not what we need
for MOND. The deep-MOND
 limit, $\mu(u)\approx u/a_0$, corresponds to
 $p=-(a_0^2/3)\varrho^3$. To get the Newtonian limit at large values
 of $u$ the equation of state has to become incompressible
 at some finite density $\varrho_0$, so that eq.~(\ref{flow}) goes
 to the Poisson equation. (Note that $p=-c^2\rho$, which is the
 relation between the energy density and pressure of the vacuum
 is not an equation of state to be applied to local distortions of
 the vacuum, which cannot be described as a fluid, in general.)
\par
 The gravitational force is then the pressure+drag force on
sources. For a small (test) static source $s$, at a position where
the fluid speed is $\vu$, the source imparts momentum to the flow
at a rate $s\vu$, and so is subject to a force $-s\gf$. The force
between sources of the same sign is attractive, as befits gravity.
The fluid density itself $\varrho$
 does not contribute to the sources of the potential equation, so
 it does not, itself, gravitate. Also note that, because
$\rho=p=0$ for $\vu=0$, the fluid behaves as if it has no
 existence without the sources (masses) that induce velocities in
 it.
This picture is still far from being directly applicable as
 an explanation of Newtonian gravity. For example, it is not clear
 how to obtain the barotropic equation of state that is needed to
 reproduce MOND. In particular, how does the infinite
 compressibility appear at a finite critical density, and what is the
 meaning of this density? Is this due to some phase transition?
 What happens at densities higher than this critical density? are
 they accessible at all? Also, there seems to be a drag force on
 moving sources, which is undesirable. Note also that in
 the context of a time-dependent configuration
 the above equation of state is problematic as it implies waves
 carrying negative energy.

 \subsection{A membrane model for MOND}
It is also well known that
 the shape of a membrane that corresponds to an extremal area
solves a nonlinear equation of type (\ref{mondpoiss}), with a
vanishing right hand side: If one describes the membrane as a
hypersurface in an $(n+1)$-dimension, Euclidean space with
coordinates $x^1,...,x^n,\ff$ taking the form $\ff(x^1,...,x^n)$,
the area (volume) of the membrane is given by
 \beq A_M=\int d^nx[1+\gffs]^{1/2},  \label{mema}\eeq
 where the integral is over the
(projected) volume in the $\vx$ space, at the boundary of which
$\ff$ is dictated. If the energy function of the membrane is
 proportional to the area, $E_M=KA_M$ its minimization gives
  eq.~(\ref{mondpoiss}) with $\mu(z)\propto(1+z^2)^{-1/2}$ (and
$\rho=0$). More generally we consider membrane energy functions
that are still functionals of the membrane shape of the form

 \beq E_M=(K/2)\int d^nx F[\gffs]. \label{jters} \eeq
\par
Then think of our effective (non-relativistic) universe as a
surface in an $(n+1)$-dimensional Euclidean (or Minkowski) space
such that at a point $(\vx,\ff)$ on the surface, the $\ff$
coordinate is to be interpreted as the gravitation potential at
$\vx$. We now have to introduce gravitating masses as sources for
the potential; in the membrane picture they will be some external
agents that determine the shape of the membrane. We can do this,
breaking the isotropy of the embedding Euclidian space, by
assuming that there are $\ff$-independent forces on the membrane
acting in the $\ff$ direction. These are introduced by adding to
the energy a term of the form $a_0\int~d^nx~\rho(\vx)\ff(\vx)$. We
can think of this term as resulting from a constant acceleration
field of magnitude \a0 acting in the $\ff$ direction, that couples
linearly to some quantity $\rho$ on the membrane with the
dimensions of mass density.
 Note that the area energy function itself is just
$\int~g^{1/2}d^nx$, where $g$ is the determinant of the induced
metric, and is covariant, and in particular isotropic in the
embedding space. It is the force term that breaks the isotropy.
\par
We can define $\phi=a_0\ff$, which then has the usual dimensions
of a gravitational potential, and also define the gravitational
constant as $G=a_0^2/4\pi K$ and then write the combined energy
(or action $S=-E$) as \beq E_M=(a_0^2/8\pi G))\int d^nx
F[\gfs/a_0^2]+\int d^nx~\rho\phi. \label{koyrs} \eeq
 The shape of
the membrane $\phi(\vr)$ that minimize the total energy is
determined by eq.~(\ref{mondpoiss}).
 \par
It can then be shown \citep[e.g.][]{milconformal,milnonlin} that
the membrane produces forces on $\rho$  {\it in the lateral, $\vx$
space}. Take a (possibly finite) body made of mass distribution
$\rho$, and define the force on it as (minus) the gradient of the
energy under rigid displacements of the body. The force is then
writable as $-\int d^nx~\rho\gf$, where $\phi(\vx)$ is the shape
of the membrane as determined by the total mass distribution,
including that of the body itself. (Because of the non-linearity
of the problem we cannot calculate the force using the potential
determined by the masses other than the body.)
 These forces appear because it costs energy to rigidly displace the body, and
they are interpreted as the gravitational
 forces on masses in the $n$-dimensional world in which we
seem to live. This justifies the viewing of $\phi$ as
 the gravitational potential.
\par
Thus, a simple dynamical picture with non-interacting ``masses''
on the membrane that are subject to a uniform force field in the
$\ff$ direction, yields a complex, effective picture of
``gravitational'' interactions between masses, mediated by the
membrane. The functional form of the dependence of the membrane
energy on its shape dictates the governing equation for the
effective gravitation field.
\par
Such membrane analogs of gravity are, of course, well known, from
the simple demonstration of effective attraction between two
masses placed on an horizontal, stretched, elastic surface, to the
many recent discussions on the universe as a brane, in the
high-energy literature \citep[see, e.g.,][and references
therein]{carter}. In such attempts, it has not yet been possible to derive
the energy function from an underlying theory (e.g. string
theory), and ad hoc energies are assumed to fill the bill.  The
emphasis here is on the potential application of the membrane to
model MOND-like theories, and this is all in the choice of the
energy functional for the membrane.
\par
The theory obtained when the energy function is just the area is
not what we need to model MOND. In the limit $|\gff|\ll 1$
($|\gf|\ll a_0$),
 we can write the area as
$\approx\int~d^nx~([1+(\gff)^2/2]$, which gives the Poisson
equation, not the deep-MOND limit. (This is generic, and tends to
happen when the energy density is finite at zero $\gff$, as the
next term in the expansion is, many times, the Poissonian
$\gffs$.)
 In the opposite limit,
$|\gffs|\gg 1$, the theory becomes singular. [In this limit
$\mu(z)\rightarrow z^{-1}$, which is exactly the borderline
between ellipticity and hyperbolicity of the field equation--the
ellipticity condition being $d~\ln[\mu(z)]/d~\ln z>-1$.] Also, the
theory does not permit  concentrated masses: Applying Gauss
theorem to the field equation
 gives

\beq \int_\sigma{\gff\cdot d\vs\over[1+\gffs]^{1/2}}=8\pi GM/a_0,
\label{hutom} \eeq
    where $M$ is the
total mass in the volume whose surface is $\sigma$, from which
follows that $M\leq a_0 S/8\pi G$, where $S$ is the total area of
$\sigma$. So, there is a limit to the mass that can be put in a
volume of a given surface area. Point masses, for example, are
excluded: in three dimensions there is no solution  for a radius
below $(2GM/a_0)^{1/2}$ since
$|\gff|=\lambda/(1-\lambda^2)^{1/2}$, where $\lambda=2MG/a_0r^2$.
\par
However, this theory does include already an important feature of
MOND: It has a critical, transition acceleration--which, in the
units where the potential is a coordinate, is $a_0=1$--that
separates two regions of rather different behavior of the
gravitational field. It is likely that this critical value will
also appear in the ``cosmology'' of the model--the behavior of the
membrane as a whole. If indeed it does, it may hint at one
possible way in which, in the real world, cosmology shares with
local dynamics this critical constant. Note that although \a0
appears here as the constant $\ff$-acceleration--a role that
smacks of cosmology--this, in itself, does not establish \a0 as an
acceleration of cosmological significance. This will have to
emerge from the dynamics of the membrane, which has to consider
whether the membrane itself is coupled to the \a0 field, and what
exactly carries inertia: the membrane, the masses $\rho$, or both.
I leave the treatment of these questions to a future discussion.
\par
 We may liken
such an appearance of \a0 in gravitational dynamics to the way the
speed of light enters relativistic kinematics of bodies--e.g. in
determining the lifetime of a moving muon--although {\it qua}
speed of light it has nothing to do with these kinematics. How
does this happen? In Minkowski-type spaces, a space-time slope of
1  plays a critical role. (For historical reasons this slope has
attained the dimensions of velocity and the value of $c$). On one
hand it enters the kinematics of all particles; on the other, it
is the constant slope on the world lines--null geodesics--of
massless particles. Similarly, in the present context, an $x-\ff$
slope of 1 for the membrane is a transition value because of the
form of the energy function of the membrane, so it enters local
dynamics, and may enter cosmology. (For historical reasons having
to do with how we measure the gravitational potential this slope
has attained the dimensions of acceleration and the value of
$a_0$.)
\par
To write a membrane energy function that does give MOND we have to
invoke again the special role of the $\ff$ direction, as we did
already when we assumed that the external field acts in the $\phi$
direction. We have to break the isotropy also in the energy
function of the membrane.
\par
For example, one of many energy functionals for a 3-dimensional
membrane embedded in 4-dimensional Euclidean space that give MOND
is
 \beq E_M=(K/2)\int~d^3x \gffs\left[{\gffs\over 1+\gffs}\right]^{1/2}.
 \label{jupapa} \eeq
This can be given a geometrical meaning: If $\psi$ is the angle
between the normal to the membrane and the (positive) $\ff$
direction, then $\cos\psi=[1+\gffs]^{-1/2}$, and
$\sin\psi=[\gffs/(1+\gffs)]^{1/2}$. So, we can write the above
energy functional as

\beq E_M=(K/2)\int~d^3x ~tg^2\psi~\sin\psi, \label{jupkaka} \eeq or
 \beq E_M=(K/2)\int~dv~ tg\psi~\sin^2\psi, \label{simna} \eeq where
$dv=d^3x/\cos\psi$ is the volume element on the membrane.
\par
We can also obtain an effective theory that lives in a curved
$n$-dimensional space by starting with a foliation of the
embedding space using coordinates $(\theta^1,...,\theta^n,\ff)$ in
which the line element can be written as \beq
ds^2=d\ff^2+\ff^2g_{ij}(\vec\theta)d\theta^id\theta^j.
\label{gomada} \eeq
 If we now describe the membrane by $\tf(\vec
\theta)$, The embedded metric on the membrane, in the coordinates
$\vec\theta$, is $G_{ij}=\tf\ud{i}\tf\ud{j}+\tf^2g_{ij}$ (where
$[]\ud{i}$ signifies the derivative with respect to $\theta_i$),
whose determinant can be shown to be
$G=g\tf^{2n}(1+g^{ij}\beta\ud{i}\beta\ud{j}) $, where $g$ is the
determinant of $g_{ij}$, $g^{ij}$ is its inverse, $\beta\equiv
\ln(\tf)$. So, the area element of the membrane is $dv=d^n\theta
G^{1/2}=
d^n\theta~g^{1/2}\tf^n~(1+g^{ij}\beta\ud{i}\beta\ud{j})^{1/2}$.
Note that the geometry of our universe is not that of the
membrane: the effective metric is not $G_{ij}$, the induced metric
on the membrane, but $g_{ij}$.
\par
 If $\psi$ is the angle between the membrane and the
local, constant-$\ff$ surface, then
$\cos\psi=(1+g^{ij}\beta\ud{i}\beta\ud{j})^{-1/2}$. Again, membrane
energy functionals that give MOND can be written that depend only
on $\psi$. An example in three dimensions is given by
expression (\ref{simna}), which makes $\ff$ a preferred direction.
It is also a preferred direction as regards the external field,
which is now assumed to lies in the $\ff$ direction--i.e. is
derived from a $\ff$-dependent potential $V(\ff)$.  Its
interaction with $\rho$, the density of masses on the membrane per
unit $d\Omega=d^3\theta$ (the covariant density being
$g^{-1/2}\rho$),
 is described by
 \beq E_I=\int~d^3\theta~\rho(\vec\theta)V[\tf(\vec\theta)].
 \label{yuter} \eeq
This would give a MOND-like theory in the curved space having the
metric $g_{ij}$ in the coordinates $\vec\theta$.
\par
Why the energy function of the model, or real, membrane should
take a form that reproduces MOND I do not know. Perhaps there is a
clue in the fact that in this limit MOND becomes conformally
invariant, as shown in
 \cite{milconformal}. To see this in the present context, note that the
form of the energy function in the deep-MOND limit
($g^{ij}\beta\ud{i}\beta\ud{j}<<1$) is
$\int~d^n\theta~g^{1/2}\tf^n~(g^{ij}\beta\ud{i}\beta\ud{j})^{3/2}$.
This is, clearly, invariant under replacement of the metric
$g_{ij}(\vec\theta)$ by $\lambda(\vec\theta)g_{ij}(\vec\theta)$,
for an arbitrary $\lambda(\vec\theta)>0$--under which
$g^{ij}\rightarrow\lambda^{-1}g^{ij}$, and $g\rightarrow
\lambda^3g$. Note that $E_I$ is also conformally invariant, so
that the field equation for $\ff$ is invariant as well.
\par
I have limited myself to the static case here, and will discuss
the dynamics of the system elsewhere.

\section{MOND as modified inertia}
 Newtonian inertia itself has not been immune to changes over the
 years. Special Relativity entails a familiar
modification, replacing the single-particle kinetic action in
eq.~(\ref{acpar})
 by $-mc^2\int~dt~[1-(v/c)^2]^{1/2}$. This gives an
equation of motion \beq
\vF=md(\gamma\vv)/dt=m\gamma[\va+\gamma^2\vv(\vv\cdot\va)/c^2],
\label{lorentz} \eeq
 where $\gamma$ is the Lorentz factor.
\par
And, physics is replete with instances of modified, acquired, or
effective inertia. Electrons and holes in solids can sometimes be
described as having a greatly modified mass tensor. Mass
renormalization and the Higgs mechanism, modify particle masses
and/or endow them with mass: an effective, approximate description
that encapsules the effects of interactions of the particles, with
vacuum fields in the former instance, and with the Higgs field in
the latter. The effects of a fluid on a body embedded in it may
sometimes be described as a contribution to the mass tensor of the
body, because its motion induces motion in the fluid which carries
energy and momentum. So, modified inertia might also well lie in
the basis of MOND.

\par
Consider non-relativistic modifications of inertia that
incorporate the basic principle of MOND. We seek to modify the
particle kinetic action $S_k$ in eq.~(\ref{acpar}) into an action
of the form $S_k[R,a_0]$, which is a functional of the particle
trajectory, $R$ [symbolically representing some trajectory
$\vr(t)$], and depends also on the single constant, \a0. The
potential part of the action remains the standard one. The
modified kinetic action should satisfy the following asymptotic
requirements: 1. In the formal limit $a_0\rightarrow 0$,
corresponding to all acceleration measures in the system being
much larger than the actual value of \a0, the action should attain
its standard Newtonian form (this is similar to obtaining the
classical limit of quantum mechanics by taking the formal limit
$\hbar\rightarrow 0$). 2. To retain MOND phenomenology, according
to which, in the deep MOND limit, $G$ and \a0 appear only through
their product $Ga_0$, we should have in the limit $a_0\rightarrow
\infty$, $S_k\propto a_0^{-1}$. This can be seen by rescaling
$\phi$ into $\phi/G$ in eq.~(\ref{acpar}) (and dividing the action
by $G$).
\par
The equation of motion is then of the form \beq
\vA(R,t,a_0)=-(\grad\phi)[\vr(t)], \label{hutsa} \eeq where the
generalized acceleration $\vA$ is a functional of the trajectory,
and a function of the time $t$, and $\grad\phi$ is
 to be calculated at the momentary position $\vr(t)$.
\par
The theory should also satisfy the  more subtle requirement of the
correct center-of-mass motion discuss in the previous section.
\par
General properties of such theories are discussed in detail in
\cite{milinertia}. Here I summarize, very succinctly, some of the
main conclusions.
\par
If the action enjoys the usual symmetries: translational,
rotational, and Galilei invariance, then, to satisfy the two
limits in \a0 the action must be non-local. This means that the
action cannot be written as $\int~L~dt$, where $L$ is a function
of a finite number of derivatives of $\vr(t)$. This might look
like a disadvantage, but, in fact, it is a blessing. A local
action for MOND would have had to be a higher-derivative theory,
and, as such, it would have suffered from the several severe
problems that beset such theories. A non-local theory need not
suffer from these.  A non-local action is also a more natural
candidate for an effective theory.
\par
While nonlocal theories tend to be rather unwieldy, they do lend
themselves to a straightforward treatment of the important issue
of rotation curves. This is done via a virial relation that
physical, bound trajectories can be shown to satisfy:

\beq 2S_k[R,a_0]-a_0{\partial S_k\over\partial a_0}=
\langle \vr\cdot\grad\phi \rangle, \label{virial} \eeq where $\phi$ is the
(unmodified) potential in which the particle is moving, $\langle \rangle$ marks
the time average over the trajectory, and $S_k$ is the value of
the action calculated for the particular trajectory ($S_k$ is
normalized to have dimensions of velocity square). In the
Newtonian case this reduces to the usual virial relation. Applying
this relation to circular orbits in an axi-symmetric potential,
and noting that, on dimensional grounds, we must have
$S_k(r,v,a_0)=v^2\nu(v^2/ra_0)$ (where $r$ and $v$ are the orbital
radius and velocity), we end up with the expression for the
velocity curve

\beq (v^2/r)\mu(v^2/ra_0)=d\phi/dr, \label{roca} \eeq
where $\mu(x)=\nu(x)[2+d\ln~\nu(x)/d\ln~x]$.
 Thus the
algebraic relation that was first used in MOND as a naive
application of eq.~(\ref{mond}), and which all existing
rotation-curve analyzes use, is exact in modified-inertia MOND. In
modified gravity this expression is only a good approximation.
\par
I recently noticed the following scaling property of deep-MOND
solutions in modified inertia: As explained above, the
single-particle kinetic action in the limit $a_0\rightarrow\infty$
has to be of the form \beq S_k(R,a_0)\approx s(R)/a_0,
\label{kopata} \eeq
 where $s$ is a functional of the trajectory
alone.
\par
In this limit, the virial relation takes the form
$s(R)=a_0 \langle \vr\cdot\gf \rangle/3$, where $s$ has the dimensions of
acceleration-times-velocity-squared.
\par
 It also follows that the equation of motion
in an external potential field becomes
 \beq \vA(R,t,a_0)=\Q(R,t)/a_0=-(\grad\phi)[\vr(t)], \label{pout} \eeq
  $\Q$ has
 dimensions of squared acceleration, and hence has the following
 scaling property: if we scale the trajectory $R$ given by
$\vr(t)$ to $R^*$ given by $\vr^*(t)=\lambda\vr(t/\zeta)$, then
 \beq \Q(R^*,t)=\lambda^2\zeta^{-4}\Q(R,t/\zeta).\label{huoy}\eeq
[The action itself scales as:
$S_k(R^*,\lambda\zeta^{-2}a_0)=\lambda^2\zeta^{-2}S_k(R,a_0)$.]
 If the potential field itself is homogeneous in $\vr$; i.e., it
  satisfies $\phi(\lambda\vr)=\lambda^{1-\alpha}\phi(\vr)$ we
  get a scaling property for the
 solutions: If $\vr(t)$ is a solution of the
 equation of motion, then so are the whole one-parameter family
 $\vr\_\zeta(t)=\zeta^{4/(2+\alpha)}\vr(t/\zeta)$.
\par
For example, the asymptotic gravitational field of a bounded mass
($\phi\propto r^{-1}$) has $\alpha=2$, so if $\vr(t)$ is a
solution, so is $\vr\_\zeta(t)= \zeta\vr(t/\zeta)$ for all
$\zeta$. The velocity on these trajectories is
$\vv\_\zeta(t)=d\vr\_\zeta/dt=\vv(t/\zeta)$, and so
 does not change with
the dilation of the orbit. This is a generalization of the notion
of asymptotic flatness of rotation curves from circular orbits to
arbitrary ones: every orbit is a member of a family of
self-similar ones, with different sizes, but
 the same velocity at the corresponding phase of the orbit.
\par
If $\phi$ is an harmonic-oscillator field (not necessarily
spherical) for which $\alpha=-1$,
 $\vr\_\zeta(t)=\zeta^4\vr(t/\zeta)$ is
always a solution in the deep-MOND regime, if $\vr(t)$ is. We see
that in this case the orbital period scales as the orbit size to
the $1/4$ power. This example is relevant, for example, for
constant-density spheres, for which the potential is harmonic.
\par
Note that for the general homogeneous (power-law) potential, the
acceleration on the trajectory $\vr\_\zeta$ is
 $\va\_\zeta=d^2\vr\_\zeta/dt^2=\zeta^{-2\alpha/(2+\alpha)}\va(t/\zeta)$, while, from the scaling of the potential field,
$(\grad\phi)(\vr\_\zeta)=\zeta^{-4\alpha/(2+\alpha)}(\grad\phi)(\vr)$.
So we see that, within the family,
 $\va\_\zeta(\vr\_\zeta)\propto[(\grad\phi)(\vr\_\zeta)]^{1/2}$.
 This proportionality evokes the basic, deep-MOND
relation eq.~(\ref{simpla}).
 But note that
it holds only when comparing corresponding points on scaled
trajectories in the same family. It is certainly not true in
general that the acceleration is proportional to the square root
of the Newtonian acceleration.

\par
This brings to mind
another important difference between the two interpretations of MOND:
In (non-relativistic) modified gravity the
 gravitational field is modified, but in this modified field
 all bodies at the same
 position undergo the same acceleration. In modified inertia the
 acceleration depends not only on position, but also on the
 trajectory. In the case of SR the acceleration depends on the
 velocity as well, but in more general theories it might depend on
 other properties of the orbit: as explained above it most probably is
a functional of the whole orbit. Because translational invariance
is retained, there is, of course,
 still a generalized momentum whose rate of change {\it is} a
 function of position only ($m\gamma\vv$ in SR) but this rate is
 not the acceleration.
So in SR, we see from eq.(\ref{lorentz}) that the acceleration for
the same force can be very different for circular
($\vv\cdot\va=0$) and for linear ($\vv\cdot\va=\pm|v||a|$)
motions.
\par
 This larger freedom in modified inertia
 comes about
 because we implement the modification via a
 modification of the action as a functional of the trajectory;
 namely,
 a function of an infinite number of variables; so,
 different trajectories might suffer different modifications.
  In modifying gravity we modify one function of the
three coordinates (the gravitational potential).
 This is an obvious point, but is worth making because in
 interpreting  data we tend to equate observed accelerations with the
 gravitational field because of our wont with Newtonian inertia.
  While this is still true in modified gravity
 it is not so in modified inertia.
 \par
 We can exemplify this point by considering the claimed anomaly in
 the motions of the Pioneer 10 and 11 spacecraft. Analysis of
 their motion have shown an unexplained effect
 \citep[see][]{anderson} that
 can be interpreted as being due to an anomalous, constant
 acceleration towards the sun of about $9\cdot 10^{-8} {\rm cm}~{\rm s}^{-2}$,
 which is of the order of \a0. This might well be due to some systematic
 error, and not to new physics. This suspicion is strengthened by
 the fact that an addition of a constant acceleration of the above
 magnitude to the solar gravitational field is inconsistent with
 the observed planetary motions (e.g. it gives a much too large
 rate of planetary perihelion precession).
MOND could naturally
 explain such an anomalous acceleration: We are dealing here with
 the strongly Newtonian limit of MOND, for which we would have to
 know the behavior of the extrapolating function $\mu(x)$ at
 $x>>1$, where $\mu\approx 1$.
  We cannot learn about this from galaxy dynamics, so we
 just parameterize $\mu$ in this region by $\mu\approx 1-\xi x^{-n}$.
 (This is not the most general form; e.g. $\mu$ may approach 1
 non analytically in $x^{-1}$, for example as $1-\exp(-\xi x)$.)
 Be that as it may, if $n=1$ we get just the desired effect in
 MOND: the acceleration in the field of the sun becomes
 $M_\odot G r^{-2}+\xi a_0$ in the sun's direction.
 In a modified gravity interpretation this would
 conflict with the observed planetary motions, which, as stated above,
 are not known to undergo such anomalous acceleration; but, in the
 modified-inertia approach it is not necessarily so. It may well
 be that the modification enters the Pioneers motion, which
 corresponds to unbound, hyperbolic motions, and the motion of
 bound, and quasi-circular trajectories in a different way. For example, the
 effective $\mu$ functions that correspond to these two motions
 might have different asymptotic powers $n$.
 If, for example, we use here tentatively
 an expression for inertia for an unbound,
 straight-line motion as given in eq.(\ref{det}) below we get in
 the Newtonian limit $F=a-c(\Lambda/3)^{1/2}$, giving exactly the effect
claimed for the Pioneer anomaly with the anomalous acceleration
$a_{an}=c(\Lambda/3)^{1/2}=7.3\cdot
10^{-8}\Omega_{\Lambda}^{1/2}(H_0/75{\rm km}~{\rm s}^{-1}{\rm
Mpc}^{-1}){\rm cm}~{\rm s}^{-2}$, with $\Omega_\Lambda\equiv
\Lambda/3H_0^2$, now believed to be around $0.7$, as deduced in
standard, non-MOND cosmology. (Note, however, that the motion of
the spacecraft is not of constant acceleration, and our Universe
is not exactly de-Sitter, as assumed in the derivation of
eq.(\ref{det}).)
\section{vacuum effects and MOND inertia}
Because MOND revolves around acceleration, which is so much in the
heart of inertia, one is directed, with the above imagery in mind,
to consider that inertia itself, not just MOND, is a derived
concept reflecting the interactions of bodies with some agent in
the background. The idea, which is as old as Newton's second law,
is the basic premise of the Mach's principle. The great sense that
this idea makes has lead many to attempt its implementation.
 The agent responsible for inertia had been taken to be
the totality of matter in the Universe.
\par
Arguably, an even better candidate for the inertia-producing
agent, which I have been considering since the early 1990s, in the
hope of understanding MOND's origin, is the vacuum. The vacuum is
known to be implicated in producing or modifying inertia; for
example, through mass renormalization effects, and through its
contribution to the free Maxwell action in the form of the
Euler-Heisenberg action \citep{euler}. Another type of vacuum
contributions to inertia have been discussed by \cite{ja}. But, it
remains moot whether the vacuum can be fully responsible for
inertia.
\par
 The vacuum is thought to be Lorentz invariant, and so indifferent to
motion with constant speed. But acceleration is another matter. As
shown by Unruh in the 1970s, an accelerated body is alive to its
acceleration with respect to the vacuum, since it finds itself
immersed in a telltale radiation, a transmogrification of the
vacuum that reflects the accelerated motion.  For an observer on a
constant-acceleration ($a$) trajectory this radiation is thermal,
with $T=\alpha a $, where $\alpha\equiv \hbar/2\pi kc$. The effect
has been also calculated approximately for highly relativistic
circular motions; the spectrum is then not exactly thermal. The
effect is non-local; i.e., depends on the full trajectory.

\par
 Unruh's
result shows that the vacuum can serve as an inertial frame. But
this is only the first step. The remaining big question is how
exactly the vacuum might endow bodies with inertia. At any rate,
what we want is the full MOND law of inertia, with the transition
occurring at accelerations of order $a_0$ that is related to
cosmology. We then have to examine the vacuum in the context of
cosmology. How it affects, and is being affected by, cosmology.
One possible way in which cosmology might enter is through the
Gibbons-Hawking effect, whereby even inertial observers in an
expanding universe find themselves embedded in a palpable
radiation field that is an incarnation of the vacuum. The problem
has been solved for de Sitter Universe, which is characterized by
a single constant: the cosmological constant, $\Lambda$, which is
also the square of the (time independent) Hubble constant. In this
case the spectrum is also thermal with a temperature $T=\alpha c
(\Lambda/3)^{1/2}$.
\par
In the context of MOND it is interesting to know what sort of
radiation an observer sees, who is accelerated in a non-trivial
universe: if the Unruh temperature is related to inertia, then it
might be revealing to learn how this temperature is affected by
cosmology. This can be gotten for the case of a
constant-acceleration observer in a de Sitter Universe. For this
case the radiation is thermal with a temperature
$T=\alpha(a^2+c^2\Lambda/3)^{1/2}$ \citep{narnhofer,deser}.
Inertia, which is related to the departure of the trajectory from
that of an inertial observer, who in de Sitter space sees a
temperature $\alpha c(\Lambda/3)^{1/2}$, might be proportional to
the temperature difference \beq \Delta
T=\alpha[(a^2+c^2\Lambda/3)^{1/2}-c(\Lambda/3)^{1/2}],
\label{det}\eeq and this behaves exactly as MOND inertia should:
it is proportional to $a$ for $a>>a_0\sim c(\Lambda/3)^{1/2}$, and
to $a^2/a_0$ for $a<<a_0$; and, we reproduce the connection of \a0
with cosmology. Of course, this sort of argument pertains only to
linear, constant-acceleration motion, while more general
trajectories will probably behave differently. But the emergence
of an expression {\it \`a-la} MOND in this connection with the
vacuum is intriguing.

\par
For inertia to result somehow from the resistance of the vacuum,
an accelerated observer should be able to tell from vacuum effects
alone, not only the magnitude of its acceleration, as in the Unruh
effect, but also all sorts
 of other properties of its orbit, as enter the generalized acceleration.
 The Unruh radiation is supposedly isotropic so its angular
distribution for a point observer does not carry directional
information. If, however, we consider a finite-size observer it is
possible that different parts of the observer are seeing different
Unruh radiation, from which difference more general properties of
the trajectory may be read. Consider, for example, two points,
infinitesimally nearby, each moving on an hyperbolic (constant
acceleration) trajectory along the $z$ axis:
$z_i(t)=(t^2+g_i^{-2})^{1/2}$, $i=1,2$. The proper distance
between the point remains constant (i.e., they move as a rigid
body) and equals $|g_2^{-1}-g_1^{-1}|$. More generally, the
different points of a rigid body, translationally accelerated at a
constant acceleration, move on parallel hyperbolic trajectories,
but each with its own acceleration parameter according to its
position along the direction of the acceleration. Each point then
has its own Unruh temperature, and the direction of the body's
acceleration can be read off the temperature distribution.

\section{Relativistic theories}
A relativistic extension of MOND, which we still do not have, is
needed for conceptual completion of the MOND idea. It is also
needed because we already have observed relativistic phenomena
that show mass discrepancies, and we must ascertain that there too
the culprit is not dark matter but modified dynamics.
\par
 There are no local
black holes whose surface acceleration is in the MOND regime;
i.e., for which $MG/r_s^2<a_0$. This would require the
Schwarzschild radius to satisfy $r_s>c^2/2a_0$, which, by the
cosmological coincidence, is larger than the Hubble distance. The
only system that is strongly general relativistic and in the MOND
regime is the Universe at large. This, however means that we would
need a relativistic extension of MOND to describe cosmology. In
fact, as I have indicated, MOND itself may derive from cosmology,
so it is possible that the question of the origin of MOND will
have to be tackled as part and parcel of that of MOND cosmology.
And, because the cosmological expansion is strongly coupled with
the process of structure formation this too will have to await a
modified relativistic dynamics for its treatment.
\par
Several relativistic theories incorporating the MOND principle
have been discussed in the literature, but none is wholly
satisfactory \citep[see, e.g.][and references therein]{bm,pcg,stratified}.
\par
There have also been attempts to supplement MOND with extra
assumptions that will enable the study of structure formation, so
as to get some glimpse of structure formation in MOND. For these
see \cite{comments}, \cite{sanstructure}, \cite{stach}, and
\cite{nusser}.
\par
Gravitational light deflection, and lensing, is another phenomenon
that requires modified relativistic dynamics. It is tempting to
take as a first approximation the deflection law of post-Newtonian
General Relativity with a potential that is the non-relativistic
MOND potential \citep[see e.g. analyses by][based on this
assumption]{qin,mortlock}. This, however, is in no
way guaranteed. In GR this is only a post-Newtonian approximation,
and perhaps it would turn out to be a post-Newtonian approximation
of MOND (i.e. an approximation of MOND in the almost Newtonian,
$a>>a_0$ regime). But, there is no reason to assume that it is
correct in the deep-MOND regime. Even in the framework of this
assumption one needs to exercise care. For example, the thin-lens
hypothesis, by which it is a good approximation to assume that all
deflecting masses are projected on the same plane perpendicular to
the line of sight, breaks down in MOND. For example, $n$ masses,
$M$, arranged along the line of sight (at inter-mass distances
larger that the impact parameter) bend light by a factor $n^{1/2}$
more than a single mass $nM$.
\par
No less important is another MOND effect without analog in
Newtonian dynamics: the external-field effect (EFE) of MOND, by
which, if a system is embedded in an external field whose
acceleration dominates over the internal acceleration, it is the
former that controls the dynamics. In particular, in the \cite{bm}
formulation of MOND as modified gravity, if a mass (say a galaxy)
is embedded in an external field (due, say, to large scale
structure) the dynamics becomes quasi-Newtonian beyond radii where
the internal acceleration falls below the external one
\citep{milsol}. Furthermore, the MOND, effective gravitational
field is non-spherical due to the EFE, even if the mass (of the
galaxy) is spherical. The field is not even elliptical,
but--assuming that the external field is constant--becomes
elliptical asymptotically with an asymptotic axes ratio of
$2^{-1/2}$. \cite{hoekstra}, for example, ignore the EFE,
altogether, when they assess the performance of MOND in light of
their galaxy, weak-lensing analysis. One of their complaints
against MOND is, in fact, that their analysis gives a ``halo
cutoff radius'' beyond which a faster, Newtonian-like fall-off in
the signal is seen, just as expected from the EFE due, e.g., to
acceleration of galaxies by LSS. This, I find, happens in the
Hoekstra et al. analysis where the galaxy's intrinsic acceleration
is $\sim 4\times 10^{-10}h~cm~sec^{-2}$, which is of the order of
the LSS acceleration of galaxies.

I thank Christophe Alard and Stephane Colombi for organizing this
workshop, and for their hospitality at the IAP.

\end{document}